\documentclass[
  aps,
  prb,
  reprint,
  superscriptaddress,
  longbibliography,
  nofootinbib,
  floatfix
]{revtex4-2}

\usepackage{amsmath,amssymb}
\usepackage{graphicx}
\usepackage[colorlinks=true,allcolors=blue]{hyperref}
\graphicspath{{figures/}}

\newcommand{\Tr}{\operatorname{Tr}}
\newcommand{\ee}{\mathrm{e}}
\newcommand{\teff}{t_{\mathrm{eff}}}
\newcommand{\RR}{\mathrm{RR}}

\begin{document}

\title{State-selective entanglement and point-gap topology in the
unidirectional Bose--Hubbard chain}

\author{Chuan-Fu Lin}
\affiliation{Department of Physics, National Tsing Hua University,
Hsinchu 30013, Taiwan}
\author{Yu-Chin Tzeng}
\affiliation{Department of ElectroPhysics and Center for Theoretical and Computational Physics, National Yang Ming Chiao Tung University, Hsinchu 300093, Taiwan}
\author{Po-Yao Chang}
\affiliation{Department of Physics, National Tsing Hua University,
Hsinchu 30013, Taiwan}

\date{September 25, 2026}

\begin{abstract}
We study the effect of repulsive interactions on skin modes and
entanglement in the unidirectional Bose--Hubbard model. Although the ground state of this
integrable model undergoes a superfluid--Mott transition with
suppressed skin accumulation, the corresponding change in entanglement
depends on the selected many-body state. We compare the ground state with
the steady state, defined as the largest-imaginary-energy eigenstate in the
zero-momentum sector. For sixteen sites at unit filling, we find that
the ground-state density approaches uniform filling and its entanglement decreases
toward the Mott limit, whereas the steady states exhibit a
nonmonotonic entanglement crossover. At intermediate interaction ($U/t=10$), the
second R\'enyi entropy of the steady state develops four spatial maxima. We show that these
maxima are accounted for by competing subsystem-number sectors within a dominant configuration manifold of
four doublons and four holons, but not
by the subsystem particle-number variance alone. The dominant global
onsite-pair manifold retains the same entropy peak structure.
At strong interaction, projection onto the one-doublon--one-holon
manifold closely reproduces the full entropy profile and its two edge peaks.
Furthermore, we demonstrate that the many-body
winding about the origin remains nonzero at strong interaction; an exact two-site example shows how it becomes ill-defined in the
singular projected $U/t\to\infty$ limit.
Our results distinguish ground-state skin suppression from state-selective entanglement and
point-gap topology.
\end{abstract}

\maketitle


\section{Introduction}

Quantum entanglement provides a way to characterize quantum many-body
states beyond local observables. The subsystem-size dependence of the
entanglement entropy distinguishes different correlation structures and
gives information about quantum criticality
\cite{Eisert2010,Calabrese2004,Laflorencie2016}. In non-Hermitian systems,
the relation between entanglement and spatial localization is particularly
interesting. Asymmetric hopping can lead to the non-Hermitian skin effect,
in which an extensive set of eigenstates accumulates near a boundary
\cite{Hatano1996,Yao2018,Kunst2018, Martinez2018}. This boundary sensitivity motivates
non-Bloch descriptions of open-boundary spectra
\cite{Yokomizo2019} and is closely related to complex-energy topology
\cite{Gong2018,Bergholtz2021,Okuma2020,Zhang2020}. However, the presence of
a skin density alone does not determine the entanglement of a many-body
state.

Entanglement provides a complementary diagnosis of non-Hermitian
states. Its spectrum can reveal topological structure and symmetry
properties \cite{Herviou2019,Chang2020}, while entropy scaling can
characterize critical systems described by nonunitary conformal field
theories \cite{Chang2020,Tu2022, Hsieh2023, Li2026, chou2026}. These diagnoses depend on the chosen
state and definition of the reduced density matrix. 
Here we use the positive reduced density matrix of a normalized right eigenstate, which quantifies its conventional bipartite entanglement under the standard Hilbert-space inner product.

It has been shown that skin modes can suppress entanglement propagation.
In a nonreciprocal spinful-fermion model, the competition between directed
particle flow and coherent coupling produces a transition from volume-law
to area-law entanglement in the normalized long-time state
\cite{Kawabata2023}. In a coupled-chain Lindblad model, delocalization of
skin steady states instead changes trajectory entanglement from area-law
to logarithmic scaling \cite{Feng2024}. These examples indicate that
the relation between skin modes and entanglement depends on the physical
mechanism and the state under consideration. A natural question is how
entanglement changes when interactions suppress the skin modes.

For interacting bosons, repulsion changes both the spatial density and
the allowed occupation configurations. Previous studies have demonstrated
interaction-suppressed skin accumulation in an integrable continuum Bose
gas \cite{Mao2023}, Mott skin effects \cite{Yoshida2024}, and many-body
wave-function structures beyond a simple density description
\cite{Hamanaka2025}. At integer filling, a lattice system can approach
a Mott state with suppressed particle-number fluctuations
\cite{Fisher1989}. In this case, removing the skin density need not restore
entanglement: the interactions can suppress both localization asymmetry
and quantum fluctuations.

Bethe ansatz offers controlled examples of interacting bosons, including
the continuum Lieb--Liniger gas \cite{Lieb1963}, integrable $q$-boson
hopping models \cite{Bogoliubov1993}, and the Bose--Hubbard dimer
\cite{Links2003,Links2006}. These are special integrable models, not an
exact solution of the generic bidirectional Bose--Hubbard chain
\cite{Choy1982}. Non-Hermiticity is also compatible with Bethe
integrability: examples include the asymmetric $XXZ$ chain
\cite{Albertini1996} and the Lieb--Liniger gas with an imaginary vector
potential \cite{Mao2023}. They provide ways to study interactions and
nonreciprocity without reducing the problem to independent particles.

The unidirectional Bose--Hubbard model provides an exactly solvable lattice
setting for this question. Its Bethe equations describe a superfluid--Mott
transition near $U/t=5.83$ at unit filling, accompanied by the suppression
of ground-state skin accumulation \cite{Zheng2024}. This ground-state
transition does not directly determine the properties of a state selected
by the largest imaginary energy. The ground state minimizes the real part of the energy,
whereas normalized nonunitary evolution can select the state with the
largest imaginary part. 
When it dominates the normalized long-time evolution, we refer to this state as the steady state.
These two states are generally different. It is
therefore important to compare their entanglement rather than infer an
entanglement transition from the ground-state density alone.

In this paper, we study state-selective skin modes and entanglement in
the unidirectional Bose--Hubbard chain. We first compute Bethe roots at
$N=M=250$ and complete many-body spectra at $N=M=8$. By inserting a flux
through the closing link, we evaluate the many-body winding number of the
complete eight-site determinant \cite{Kawabata2022}. We find that the
point-gap winding about the origin remains nonzero at strong interaction, even when the
ground-state skin density is strongly suppressed. An exact two-site
example illustrates how a spectral loop can shrink without changing its
winding at finite interaction. Furthermore, we demonstrate that the nontrivial spectral monodromy of the Bethe roots from the flux insertion
 can detect the superfluid-Mott transition.

We then compare the ground state and steady state at
$N=M=16$. The ground-state entanglement decreases toward the Mott limit,
while the selected excited states show a nonmonotonic crossover and four
spatial maxima of the second R\'enyi entropy at intermediate interaction.
We demonstrate that the subsystem particle-number variance does not
reproduce all four maxima. Instead, resolving
the reduced density matrix into subsystem-number sectors
\cite{Goldstein2018,Xavier2018} identifies a competition between neighboring integer
charges. This separation of number and configurational contributions
also gives a physical interpretation of entanglement in interacting
bosonic systems \cite{Lukin2019}. At intermediate interaction, the maxima of the second Ren\'yi entropy arise from
the four-doublon--four-holon manifold, while at strong interaction, the two maxima of the second Ren\'yi entropy located at the edges are accounting for by the 
one-doublon--one-holon manifold.
The contrast with the ground state is
the central result of this work.

The paper is organized as follows. Section~II introduces the model, Bethe
roots, and many-body winding number. Section~III defines the observables
and state selections. Section~IV compares ground-state and
steady-state entanglement and particle-number fluctuations,
including the particle-number-sector and global onsite-pair analyses.
Section~V examines selected low-real-energy states, and Sec.~VI concludes.

\section{Model, Bethe roots, and the many-body winding number}
\begin{figure*}[!tp]
 \centering
 \includegraphics[width=\textwidth]
 {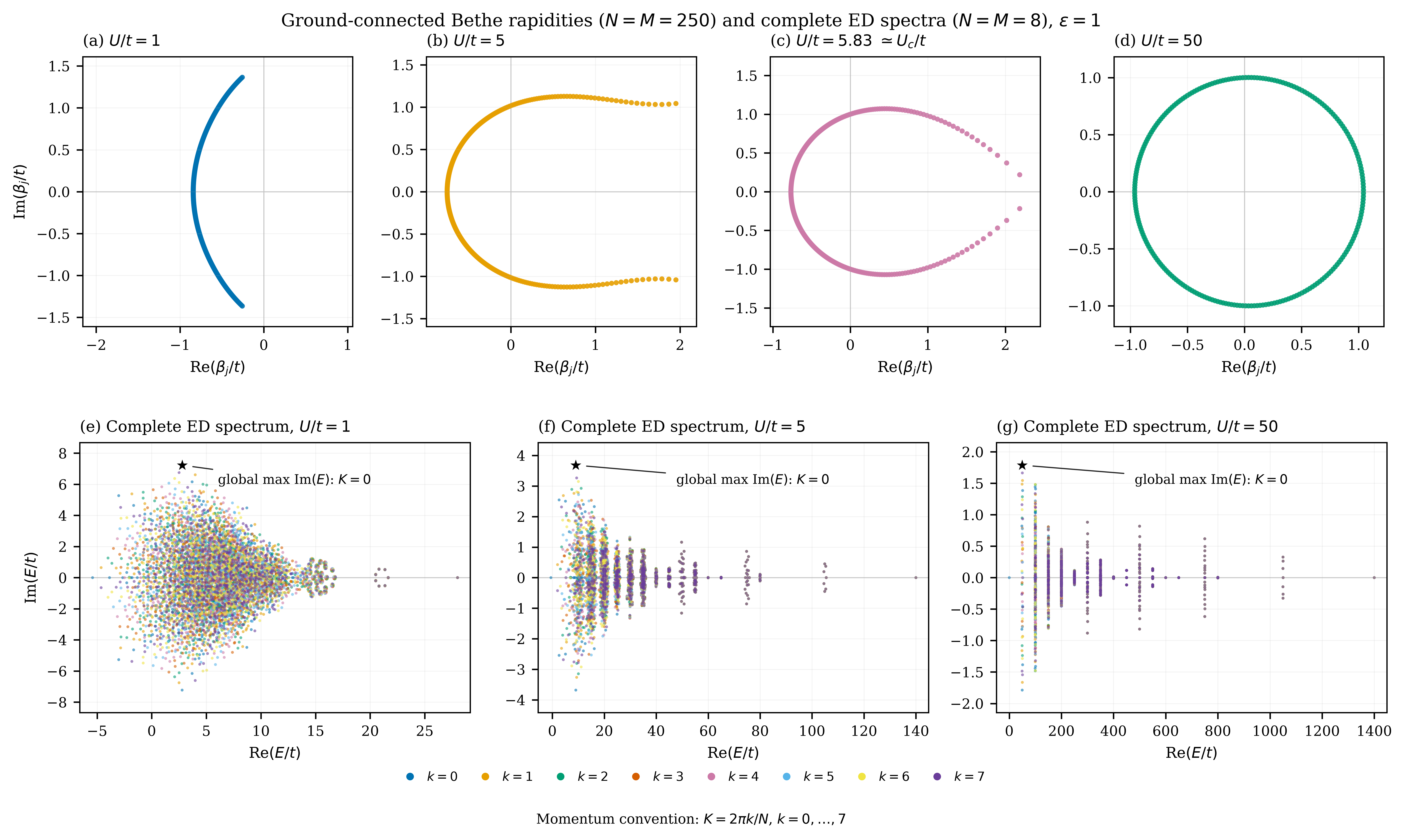}
 \caption{Large-system Bethe roots and complete finite-system spectra at
 $\epsilon=1$.  Top: the 250 roots $\beta_j/t$ of the
 ground-connected branch at $N=M=250$ and $U/t=1,5,5.83,50$. 
Bottom: all many-body energies from exact diagonalization
 at $N=M=8$ for $U/t=1,5,50$; colors label the integer momentum index
 $k$, with $K=2\pi k/N$. Black stars mark the unique global
 largest-$\operatorname{Im}E$ eigenvalue in each spectrum. All three
 marked states belong to $K=0$.
 }
 \label{fig:betheED}
\end{figure*}

We consider $M$ bosons on an $N$-site chain,
\begin{equation}
 H=-t\left(\sum_{j=1}^{N-1}b_j^\dagger b_{j+1}
 +\epsilon b_N^\dagger b_1\right)
 +\frac{U}{2}\sum_{j=1}^{N}n_j(n_j-1),
 \label{eq:H}
\end{equation}
where hopping is unidirectional from $j+1$ to $j$, including from site $1$ to
site $N$ across the closing link. The link is weak for $|\epsilon|<1$. We work at unit filling $M=N$ and set $t=1$
unless stated otherwise. The fixed-$M$ Hilbert-space dimension is
$\mathcal D_{N,M}=\binom{N+M-1}{M}$.

For real positive $\epsilon$, define
\begin{equation}
 \eta=\epsilon^{1/N},\qquad
 S=\exp\!\left[(\ln\eta)\sum_{j=1}^{N}(j-1)n_j\right].
 \label{eq:S}
\end{equation}
The exact local similarity relation \cite{Zheng2024} is
\begin{equation}
 H(t,\epsilon,U)=S H(t\epsilon^{1/N},1,U)S^{-1}.
 \label{eq:similarity}
\end{equation}
Thus the spectrum depends on the boundary parameter through
$\teff=t\epsilon^{1/N}$, while the right eigenvector is spatially
reweighted by $S$. The uniform-link Hamiltonian on the right-hand side of
Eq.~\eqref{eq:similarity} is translation invariant. Thus, we can define $K$ as its total momentum. 
Although the weak-link Hamiltonian at $|\epsilon|\ne1$
is not translation invariant, its right eigenstate follows from the
uniform-link eigenstate $|R_0\rangle$ as
\begin{equation}
 |R\rangle=\frac{S|R_0\rangle}{\|S|R_0\rangle\|}.
 \label{eq:physicalRightState}
\end{equation}

An algebraic Bethe ansatz solves this model~\cite{Zheng2024}
within the quantum inverse scattering framework
\cite{Korepin1993,Links2003}.
The transfer-matrix eigenvalues satisfy
\begin{align}
 \Lambda(z)={}&z^N\prod_{j=1}^{M}
 \frac{z-\beta_j/U-1}{z-\beta_j/U}\nonumber\\
 &+\epsilon\left(-\frac{t}{U}\right)^N
 \prod_{j=1}^{M}\frac{z-\beta_j/U+1}{z-\beta_j/U}.
 \label{eq:TQ}
\end{align}
Cancellation of the apparent poles yields
\begin{equation}
 \left(\frac{\beta_j}{-t}\right)^N
 =\epsilon\prod_{\ell\ne j}
 \frac{\beta_j-\beta_\ell+U}{\beta_j-\beta_\ell-U},
 \qquad j=1,\ldots,M.
 \label{eq:Bethe}
\end{equation}
The many-body energy is
\begin{equation}
 E=\sum_{j=1}^{M}\beta_j.
 \label{eq:BetheEnergy}
\end{equation}
Thus the $\beta_j$ are interacting Bethe roots.  Only at
$U=0$ can they be interpreted as independent one-particle energies; for
$U>0$, every root depends collectively on the other $M-1$ roots.  In
particular, a set of $M$ roots produces one many-body eigenvalue through
Eq.~\eqref{eq:BetheEnergy}.

The top row of Fig.~\ref{fig:betheED} shows the Bethe roots for $N=M=250$
from weak coupling $U/t=1$, intermediate coupling $U/t=5, 5.83$, and strong coupling $U/t=50$.
It was shown that the thermodynamic transition is near
$(U/t)_c\simeq5.83$ \cite{Zheng2024}, where the limiting root-support
curve closes near the positive real axis; the finite-size roots remain distinct.
I.e., the open arc has nearly pinched into a closed loop. 
This finite-$N$ morphology
is consistent with the reported transition, but it is not an independent
critical-point extraction or a thermodynamic scaling analysis.  

\begin{figure*}[!tp]
 \centering
 \includegraphics[width=\textwidth]{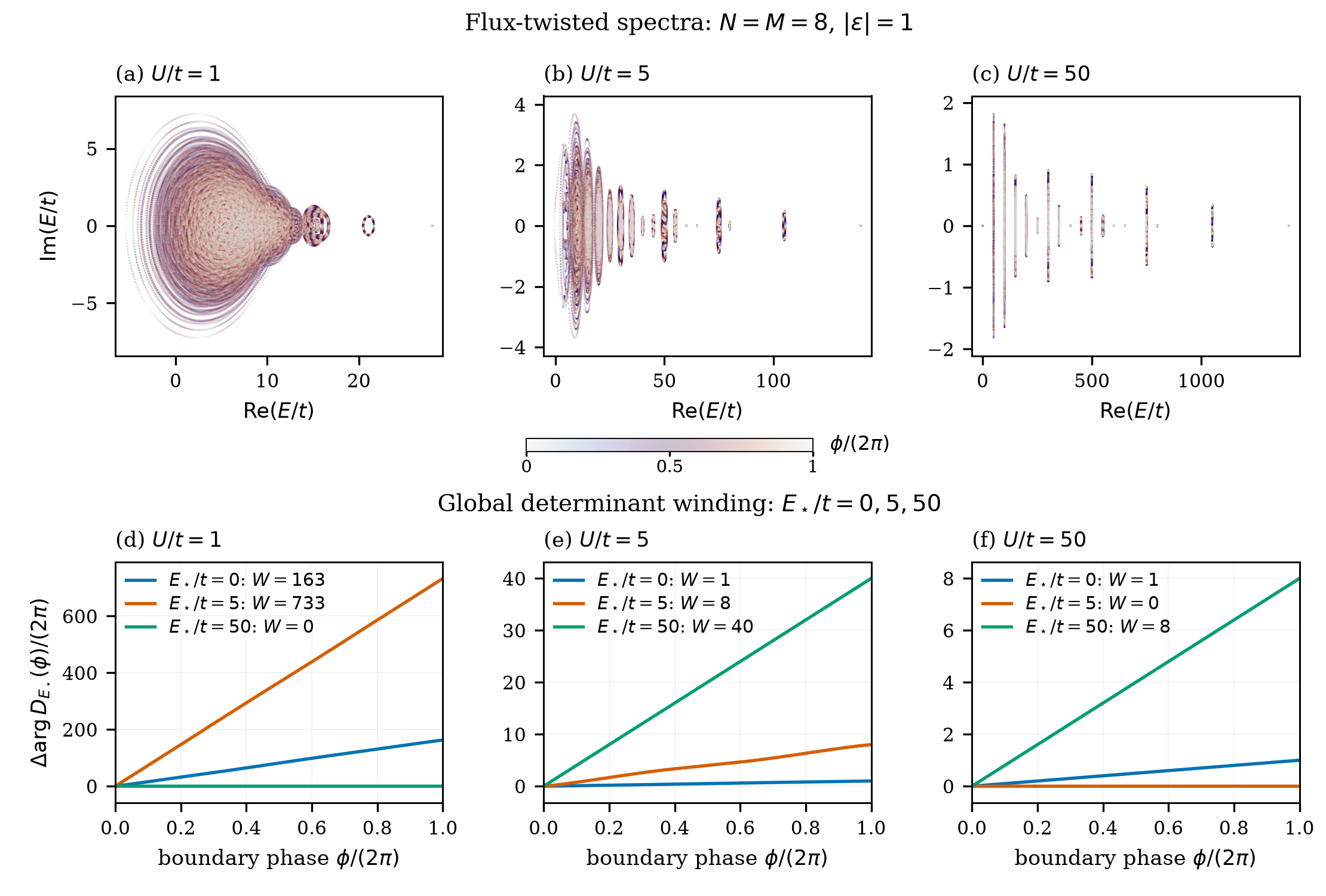}
 \caption{Flux-twisted many-body spectra and determinant winding with
 $N=M=8$ and $\epsilon(\phi)=\ee^{i\phi}$. Columns correspond to
 $U/t=1,5,50$ in both rows. (a)--(c) Complete complex spectra, including
 all eight momentum blocks of the uniform-flux representative. Color
 denotes the closing-link phase: $\phi_j=2\pi j/31$,
 $j=0,\ldots,30$, with 6435 energies per phase. 
 (d)--(f) Determinant phase evolution for the same systems at three
 real reference energies, $E_\star/t=0,5,50$. The vertical axis is the
 change of the unwrapped phase of
 $\mathcal D_{E_\star}(\phi)=\det[H(\phi)-E_\star]$, measured relative
 to $\phi=0$ and divided by $2\pi$.}
 \label{fig:windingN8}
\end{figure*}

Although solving Eq.~\eqref{eq:Bethe} for one chosen branch
makes selected large-system states accessible, 
the complete enumeration would still require an astronomical number
of distinct sheet-resolved root sets, including singular
and coalescent solutions. Therefore, in the following discussion,
we focus on the exact sector diagonalization.
The lower row of Fig.~\ref{fig:betheED} gives
the complete comparison at $N=M=8$.

The starred states in Fig.~\ref{fig:betheED}(e)--(g),
 which have the largest imaginary part of the energy, belong to $K=0$. 
For the uniform representative, an initial state in $K=0$ remains in
$K=0$ under the dynamics. 
A unique largest-imaginary-energy eigenstate in that subspace governs
the long-time limit when its initial spectral weight is nonzero.
In the following, 
we refer to the state selected by Eq.~\eqref{eq:maxamp} as the steady state.

For $|\epsilon|<1$, the model exhibits skin accumulation that is strongly
suppressed in the ground state at large $U/t$ \cite{Zheng2024}. To compare
this density suppression with spectral topology, we compute the
many-body point-gap winding number by inserting a flux
\cite{Gong2018,Kawabata2022}.
For a flux-twisted closing link $r\ee^{i\phi}$, we define
\begin{equation}
 W(E_\star)=\frac{1}{2\pi i}\int_0^{2\pi}d\phi\,
 \partial_\phi\ln\det[H(\phi)-E_\star].
 \label{eq:winding}
\end{equation}
This winding number characterizes the complete determinant, not the
ground-state eigenvalue alone.

\paragraph{Flux continuation of the ground-connected Bethe roots.}
The root morphology in Fig.~\ref{fig:betheED} motivates a direct
connection between the Bethe solution and flux response. We replace
$\epsilon$ by $\ee^{i\phi}$ in Eq.~\eqref{eq:Bethe} and continuously
follow the root set that starts at the ground state at $\phi=0$,
without reselecting the minimum-$\operatorname{Re}E$ state at each
phase. Its energy is
\begin{equation}
 E_g^{\rm cont}(\phi)=\sum_{j=1}^{M}\beta_j(\phi).
 \label{eq:BetheFluxEnergy}
\end{equation}
For $N=M=250$, the branches at $U/t=1$ and $5$ reach different
many-body energies after a single flux quantum. We find
$\Delta E_g=E_g^{\rm cont}(2\pi)-E_g^{\rm cont}(0) \neq 0$.
Since the physical Hamiltonian returns at $2\pi$, these unequal
energies demonstrate nontrivial spectral monodromy.
I.e., the continuation takes the initial eigenvalue to another eigenvalue of the same
Hamiltonian. A single open segment of such a permutation cycle
does not have an integer single-state winding.

At $U/t=6$, above the reported thermodynamic transition, the sum of the
root set remains unchanged  after one flux quantum, i.e., $\Delta E_g=E_g^{\rm cont}(2\pi)-E_g^{\rm cont}(0)= 0$.  
These results connect the open-arc and closed-ring root regimes
to different flux responses at representative couplings on the two
sides of the transition~\cite{Zheng2024}. 


Finally, Eq.~\eqref{eq:winding} involves all eigenvalues, not just
Eq.~\eqref{eq:BetheFluxEnergy}. Formally, a complete, nonredundant
enumeration of physical Bethe solutions $\alpha$, with algebraic
multiplicities and appropriate singular limits where necessary, gives
\begin{equation}
 \det[H(\phi)-E_\star]
 =\prod_{\alpha=1}^{\mathcal D_{N,M}}
 \left[\sum_j\beta_j^{(\alpha)}(\phi)-E_\star\right].
 \label{eq:BetheDeterminant}
\end{equation}
Individual eigenvalues can permute under flux, while their product
is single valued. The large-system Bethe continuation thus supplies
a state-resolved connection to spectral topology. Here we evaluate the
full finite-size determinant directly in all translation sectors,
without assuming complete enumeration of Bethe solutions; this winding
calculation itself does not require diagonalization.

Figs.~\ref{fig:windingN8}(a)--(c) show the spectral response to
the twist using the complete $N=M=8$ spectrum at each displayed phase.
For $|\epsilon|=1$, a unitary gauge transformation spreads the boundary
phase uniformly, giving hopping $t\ee^{i\phi/N}$ and preserving the
eight translation blocks. The reference dependence distinguishes different point-gap regions.
The nonzero origin winding at $U/t=50$ shows that strong ground-state density
suppression need not coincide with loss of the complete-spectrum point
gap. A two-site example makes the distinction explicit: the winding
remains nonzero for every finite positive interaction, and becomes
undefined when its loop collapses onto the reference energy.

For the two-site system $N=M=2$, the Hamiltonian in the normalized
basis $(|20\rangle,|11\rangle,|02\rangle)$ is
\begin{equation}
 H_2(\phi)=
 \begin{pmatrix}
 U&-\sqrt2t&0\\
 -\sqrt2t\,r\ee^{i\phi}&0&-\sqrt2t\\
 0&-\sqrt2t\,r\ee^{i\phi}&U
 \end{pmatrix}.
 \label{eq:H2}
\end{equation}
Its eigenvalues and determinant are
\begin{align}
 E_0&=U,\qquad
 E_\pm=\frac{U\pm\sqrt{U^2+16rt^2\ee^{i\phi}}}{2},\nonumber\\
 \det H_2(\phi)&=-4Urt^2\ee^{i\phi}.
 \label{eq:N2spectrum}
\end{align}
Consequently $W(0)=1$ for every finite $U>0$ and $r>0$. At strong coupling,
\begin{equation}
 E_-(\phi)=-\frac{4rt^2}{U}\ee^{i\phi}+O(U^{-3}),
 \label{eq:N2loop}
\end{equation}
so the loop radius vanishes as $U^{-1}$ while the finite-$U$ winding remains
one. Only in the singular projected $U\rightarrow\infty$ limit does the loop
collapse onto $E_\star=0$, where the point gap closes and $W(0)$ becomes
undefined rather than continuously becoming zero.

\section{Observables and entanglement}

To study the properties of these skin modes under different $U/t$,
we consider several quantities including the  Ren\'yi entropies.
We consider right states satisfying $H|R_m\rangle=E_m|R_m\rangle$ and  
normalized as $\langle R_m|R_m\rangle=1$.

First, we compute the local particle number for each state 
\begin{equation}
\langle n_j\rangle_{\RR, m} = \langle R_m| b^\dagger_j b_j |R_m\rangle.
\end{equation}
This quantity direct reveals skin modes, which accumulate at the left edge.
To further quantify the localization property of the skin modes,  we define the right-state density asymmetry by the half-chain
imbalance
\begin{equation}
 \delta_n=\frac{1}{M}\left|
 \sum_{j=1}^{N/2}\langle n_j\rangle_{\RR}
 -\sum_{j=N/2+1}^{N}\langle n_j\rangle_{\RR}\right|.
 \label{eq:diagnostics}
\end{equation}

In terms of the entanglement quantities,
for a subsystem of length $\ell_A$ on the left, $A=\{1,\ldots,\ell_A\}$, we define the reduced density matrix
\begin{equation}
 \rho^{\RR}_{A,m}=\Tr_B|R_m\rangle\langle R_m|.
 \label{eq:rdms}
\end{equation}
The R\'enyi entropies are
\begin{align}
 S^{\RR}_{n,m}&=\frac{1}{1-n}\ln\Tr[(\rho^{\RR}_{A,m})^n],
 \quad n>1.
 \label{eq:RRentropies}
\end{align}
For simplicity, we consider only the second  R\'enyi entropy.

To motivate the steady-state selection, consider normalized Hamiltonian
evolution,
\begin{equation}
 |\psi(\tau)\rangle=
 \frac{\ee^{-iH\tau}|\psi(0)\rangle}
 {\|\ee^{-iH\tau}|\psi(0)\rangle\|}.
 \label{eq:normalizedDynamics}
\end{equation}
If there is a unique eigenvalue with the largest imaginary part and a
nonzero initial spectral weight, the long-time state is its right eigenvector
up to an overall phase. This state need not minimize the real energy.
For the numerical steady-state search, we restrict the uniform
representative to $K=0$ and select
\begin{equation}
 m_\star^{(K=0)}=
 \operatorname*{arg\,max}_{m\in\mathcal C_{K=0}}\operatorname{Im}E_m,
 \label{eq:maxamp}
\end{equation}
where $\mathcal C_{K=0}$ is the finite list of eigenstates in the zero-momentum sector.

The ground-state index is selected as
\begin{equation}
 g=\operatorname*{arg\,min}_m\operatorname{Re}E_m.
 \label{eq:selections}
\end{equation}

For further comparison, we consider the connected number correlators, 
\begin{align}
 C_{nn}(j,k)&=\langle n_jn_k\rangle_{\RR}
 -\langle n_j\rangle_{\RR}\langle n_k\rangle_{\RR}.
 \label{eq:variance}
\end{align}
For subsystem length $\ell_A$, the subsystem contains
$N_A=\sum_{j=1}^{\ell_A}n_j$ particles. Its variance is
\begin{align}
 \operatorname{Var}_{\RR}(N_A)
 &=\langle N_A^2\rangle_{\RR}-\langle N_A\rangle_{\RR}^2\nonumber\\
 &=\sum_{i,j\in A}C_{nn}(i,j)
 =-\sum_{i\in A,j\in B}C_{nn}(i,j).
 \label{eq:crossvariance}
\end{align}
The last equality follows from fixed total particle number. This
subsystem variance includes correlations among all sites in $A$ and
is distinct from the onsite variance $C_{nn}(j,j)$.

\section{State-selective skin modes and their entanglement}
We focus on $N=M=16$ and $\epsilon=0.1$. We first study the ground
state and then the steady state from the state-search with the largest imaginary energy selected by
Eq.~\eqref{eq:maxamp} in the zero-momentum sector. 

\subsection{Ground-state entanglement for sixteen sites}
\begin{figure*}[!tp]
 \centering
 \includegraphics[width=\textwidth]{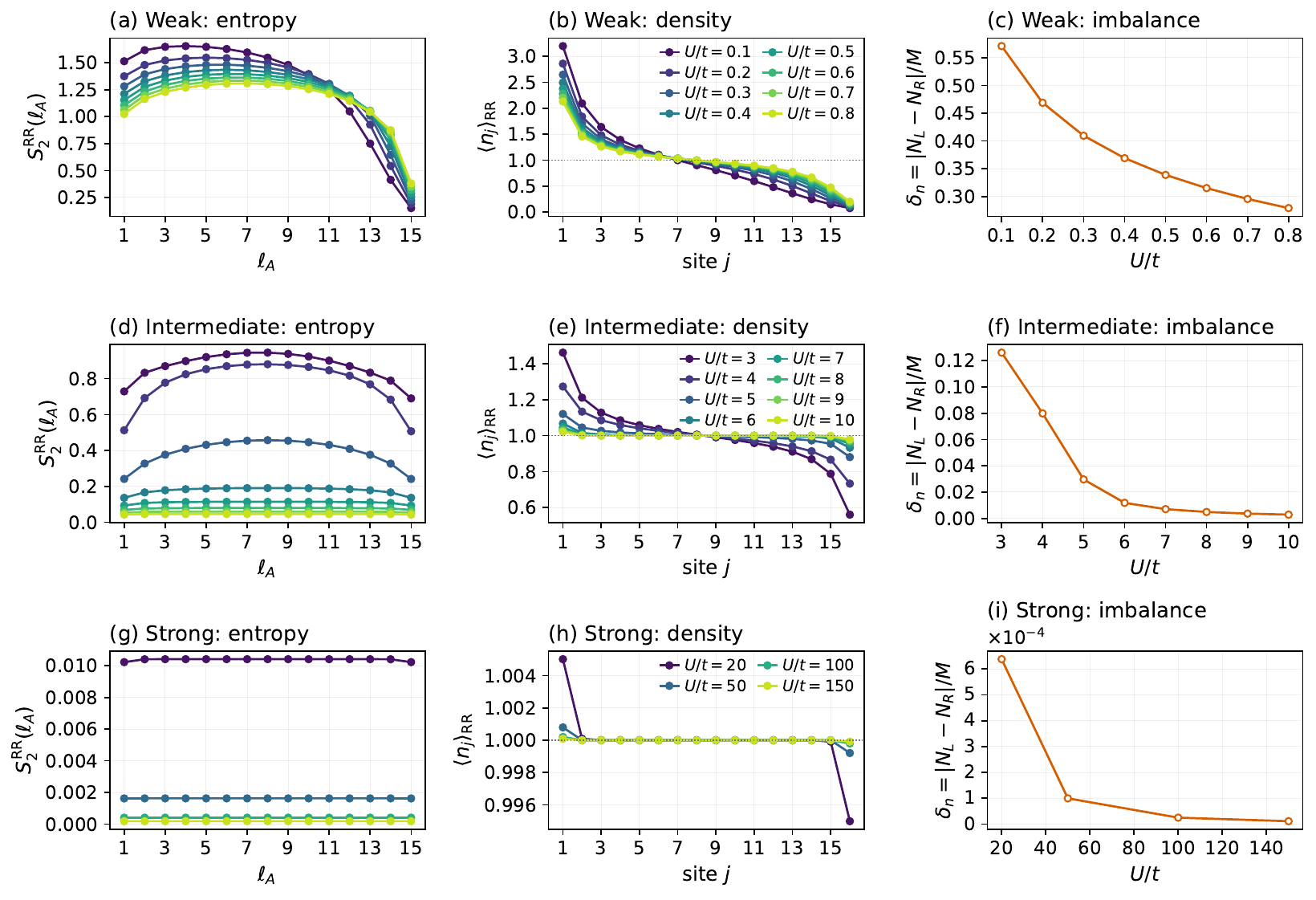}
 \caption{Second Ren\'yi entropy and density of the ground state for $N=M=16$ and $\epsilon=0.1$.
 Rows show the same parameter grids as
 Fig.~\ref{fig:amplificationN16}: (a)--(c) weak coupling,
 $U/t=0.1,0.2,\ldots,0.8$; (d)--(f) intermediate coupling,
 $U/t=3,4,\ldots,10$; and (g)--(i) strong coupling,
 $U/t=20,50,100,150$.  The left column gives second Ren\'yi entropy as a function of subsystem size $\ell_A$, $\ell_A=1,\ldots,15$.  
 The middle column gives the physical density, and the right column gives the
 half-chain imbalance $\delta_n$ versus $U/t$.}
 \label{fig:groundN16}
\end{figure*}

Fig.~\ref{fig:groundN16} shows the ground-state results for
 $N=M=16$ and $\epsilon=0.1$ in all three interaction regimes. At weak interaction ($U/t=0.1$ to $0.8$),
the second Ren\'yi entropy at the left edge decreases
 which has the same trend as the suppression of the skin accumulation at the left edges.
The maximum of the second Ren\'yi entrop moves toward the center as $U/t$ increases. 
At intermediate interaction ($U/t=3$ to $10$), while the skin mode gets strongly suppressed,
the second Ren\'yi entropy become more reflection symmetric which the  half-chain imbalance $\delta_n$ also goes to zero.
At strong interaction ($U/t=20$ to $150$),
the ground state approaches to zero entanglement, as expected for a Mott state.

This behavior differs qualitatively from steady state. The ground state loses both density accumulation and
second Ren\'yi entropy consistent with the strong-coupling Mott limit. On the other hand, the
steady state at strong-interaction limit instead retains an asymmetric density
and a different cut-entropy profile as discussed below.

\subsection{Steady state entanglement for sixteen sites}

\begin{figure*}[!tp]
 \centering
 \includegraphics[width=\textwidth]{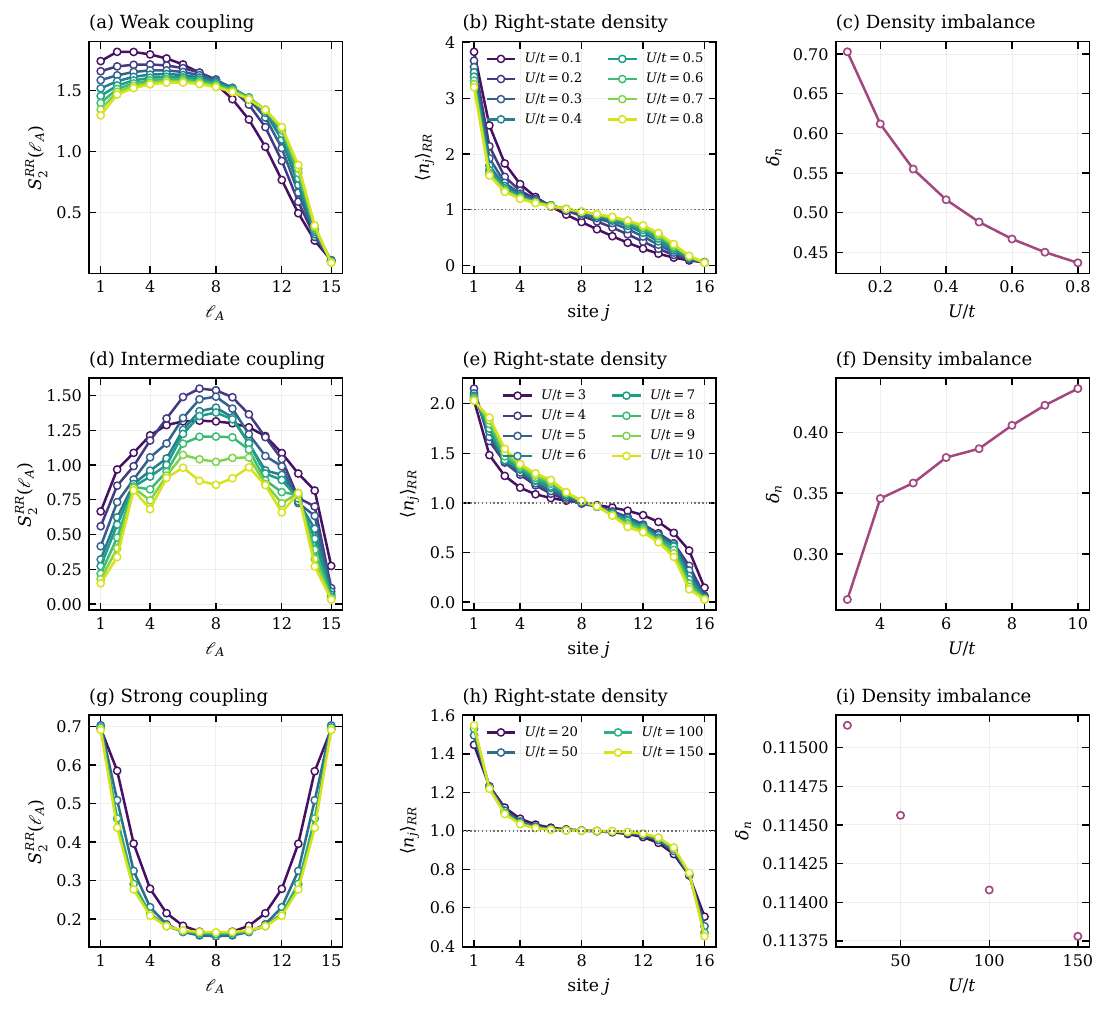}
 \caption{Second Ren\'yi entropy and density of the steady state for $N=M=16$
 and $\epsilon=0.1$. Rows show (a)--(c) weak coupling,
 $U/t=0.1,0.2,\ldots,0.8$; (d)--(f) intermediate coupling,
 $U/t=3,4,\ldots,10$; and (g)--(i) strong coupling,
 $U/t=20,50,100,150$. The left column gives the second Ren\'yi entropy as a function of $\ell_A$, $\ell_A=1,\ldots,15$.
 The middle column shows the normalized physical right-state density. The right column gives the half-chain imbalance $\delta_n$ of
 Eq.~\eqref{eq:diagnostics}. }
 \label{fig:amplificationN16}
\end{figure*}

\subsubsection{Weak interaction}
At $U=0$, the single-particle problem is the unidirectional limit of
the Hatano--Nelson model \cite{Hatano1996}. For bosons, however, the
entanglement of a condensate is different from that of the free-fermion
states studied in Ref.~\cite{Kawabata2023}. For even $N$ at unit filling $M=N$, a chosen maximizing
single-particle skin orbital gives a bosonic condensate.  For the half cut
$\ell_A=N/2$ at fixed $0<\epsilon<1$, its subsystem weight is
$p_A=1/(1+\epsilon)$ and the Schmidt probabilities are
\begin{equation}
 p_n=\binom{N}{n}p_A^n(1-p_A)^{N-n}.
 \label{eq:binomial}
\end{equation}
Hence
\begin{equation}
 S_2^{\RR}(N/2)=\frac12\ln[4\pi Np_A(1-p_A)]+o(1).
 \label{eq:noninteracting}
\end{equation}
Fixed-$\epsilon$ scale-free skin localization therefore produces logarithmic
number entanglement, not a strict area law.  The distinction between
scale-free and fixed-localization-length limits is important
\cite{Fu2023}; the fermionic volume-to-area scenario
\cite{Kawabata2023} cannot simply be transferred to this condensate. 

Figs.~\ref{fig:amplificationN16}(a)--(c) show the weak interaction results.
The second Ren\'yi entropy has no reflection symmetry due to the skin effect. 
The right-state density shows the suppression of the skin effect as $U/t$ increases,
as reflected in the decrease of density imbalance $\delta_n$. This decrease
indicates reduced skin effect for the steady state.

\begin{figure*}[!tp]
 \includegraphics[width=\textwidth]{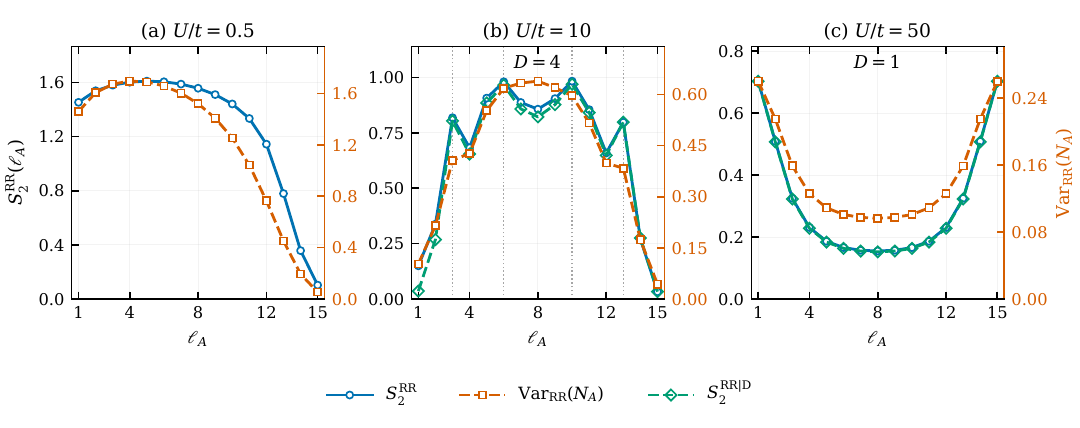}
 \caption{Particle-number fluctuations and projected entropies for the
 $N=M=16$, $\epsilon=0.1$, $K=0$
 state-search with the largest imaginary energy at (a) $U/t=0.5$, (b) $U/t=10$,
 and (c) $U/t=50$. Blue circles show $S_2^{\RR}(\ell_A)$ on the left axis;
 orange squares show the subsystem variance
 $\operatorname{Var}_{\RR}(N_A)$ of Eq.~\eqref{eq:crossvariance}
 on the right axis. Panels (b) and (c) additionally show the second
 R\'enyi entanglement $S_2^{\RR|D}$ of the normalized global-$D$-projected
 state [Eq.~\eqref{eq:Dprojectedentropy}, green diamonds] for $D=4$ and
 $D=1$, respectively. All entropies use natural logarithms. Separate
 axes retain the physical amplitudes; no curve is normalized to its
 maximum. Dotted lines in panel (b) mark the four strict $S_2^{\RR}$ maxima at
 $\ell_A=3,6,10,13$, whereas the subsystem variance peaks only at $\ell_A=8$.
 The $D=4$ projection retains those four maxima, and the $D=1$
 projection retains the two endpoint maxima at $\ell_A=1,15$ in panel (c).
 The weights $w_4=0.79036$ and $w_1=0.98954$ are state probabilities,
 not additive entropy fractions. The $U/t=50$ state is a retained
 excited branch, not a certified global maximal-$\operatorname{Im}E$ state.}
 \label{fig:mechanism}
\end{figure*}

\subsubsection{Intermediate interaction}
\label{sec:intermediate}

Figs.~\ref{fig:amplificationN16}(d)--(f) present the intermediate
interaction results. 
We observe the development of several spatial maxima of the second R\'enyi entropy.
On the other hand, the density imbalance increases when the interaction strength increases at this intermediate
interaction. This feature is opposite to the weak-interaction and the strong-interaction trends.

We specifically focus at $U/t=10$ where $S_2^{\RR}$ has strict maxima at $\ell_A=3,6,10,13$.
Fig.~\ref{fig:mechanism}(b) compares it directly with the subsystem
particle-number variance. Although both quantities are enhanced in the
interior, $\operatorname{Var}_{\RR}(N_A)$ has only one broad maximum
at $\ell_A=8$. The four entropy peaks
therefore cannot be inferred from the variance, which retains only the
second cumulant of the subsystem-number distribution.

We now resolve the origin of the four spatial maxima of $S_2^{\RR}$ for the steady state of $N=M=16$, $\epsilon=0.1$, $U/t=10$. 
We identify the dominant global onsite-pair manifold and examine the
number distribution after projection onto it. We find that the projected entanglement of the dominant global onsite-pair manifold follows 
the four spatial maxima of $S_2^{\RR}$.

\paragraph{Global onsite-pair manifolds.}
\label{sec:globalD}

To identify which occupation configurations support this pattern,
define the global onsite-pair operator
\begin{equation}
 \widehat D_{\rm tot}=
 \sum_{j=1}^{16}\frac{\widehat n_j(\widehat n_j-1)}{2},
 \qquad H_{\rm int}=U\widehat D_{\rm tot}.
 \label{eq:Dhat}
\end{equation}
It counts pairs of bosons occupying the same site across
the entire chain, not the particles in $A$. In the occupation basis,
\begin{equation}
 D(\boldsymbol n)=\sum_{j=1}^{16}\binom{n_j}{2},
 \qquad \sum_{j=1}^{16}n_j=16.
 \label{eq:Dfock}
\end{equation}
A site with zero or one boson contributes no pair; occupations two,
three, and four contribute one, three, and six pairs, respectively.
Thus the global $D=4$ manifold is
\begin{equation}
 \mathcal H_4=\operatorname{span}
 \{|\boldsymbol n\rangle:\ \textstyle\sum_jn_j=16,\,
 D(\boldsymbol n)=4\}.
 \label{eq:D4space}
\end{equation}

There are two allowed occupation compositions for $ D(\boldsymbol n)=4$.
One contains four doublons and four holons; the other contains one
triplon, one doublon, and three holons.
We list these two compositions in Table~\ref{Table1}.

\begin{table*}
\centering
\begin{tabular}{c|r|r|r|r}
 & Sites with $3$ bosons & Sites with $2$ bosons & Sites with $1$ boson & Empty sites\\
\hline
Four doublons and four holons & 0 & 4 & 8 & 4\\
One triplon, one doublon, and three holons & 1 & 1 & 11 & 3
\end{tabular}
\caption{Two allowed occupation compositions for $ D(\boldsymbol n)=4$.}
\label{Table1}
\end{table*}

Both compositions contain sixteen particles and four onsite pairs in $D=4$ manifold.
No site with $n\geq4$ is allowed in $\mathcal H_4$, since it would
already contribute at least six pairs.
Direct counting in the physical $U/t=10$ right state gives conditional
probabilities $0.998388$ and $0.001612$ for the four-doublon pattern
and the triplon--doublon pattern, respectively. Thus this particular
$D=4$ component is dominated by
four doublons and four holons. 

For a normalized physical right state $|R\rangle$, define the
global projector, its probability weight, and the normalized
conditional state for a general pair count $D$:
\begin{align}
 \Pi_D&=\sum_{\boldsymbol n:D(\boldsymbol n)=D}
 |\boldsymbol n\rangle\langle\boldsymbol n|,\nonumber\\
 w_D&=\langle R|\Pi_D|R\rangle,\qquad
 |R_D\rangle=\frac{\Pi_D|R\rangle}{\sqrt{w_D}}.
 \label{eq:D4projector}
\end{align}
The projector is independent of the cut and preserves all sixteen
particles. Hopping generally changes $D$, so $D$ is not a conserved
symmetry quantum number of the full Hamiltonian. This projection is
therefore distinct from particle-number symmetry resolution.
The actual second R\'enyi entanglement of the projected state is
\begin{align}
 \rho_A^{\RR|D}&=\Tr_B|R_D\rangle\langle R_D|,\nonumber\\
 S_2^{\RR|D}(\ell_A)&=-\ln\Tr[(\rho_A^{\RR|D})^2].
 \label{eq:Dprojectedentropy}
\end{align}

Its conditional subsystem-number probabilities are
\begin{equation}
 p_q^{(D)}(\ell_A)
 =\frac{\langle R|\Pi_D\Pi_{N_A(\ell_A)=q}|R\rangle}{w_D}.
 \label{eq:D4conditional}
\end{equation}
Here $\Pi_{N_A(\ell_A)=q}$ selects $q$ particles in $A$ and therefore
$16-q$ particles in $B$. This is independent of the onsite-pair
condition $D_A+D_B=D$; in particular, $D=4$ does not impose four
particles in either subsystem. The two projectors commute because
both are diagonal in the occupation basis. For $U/t=10$, the
$D=4$ projected entanglement $S_2^{\RR|4}$ retains all four maxima of
$S_2^{\RR}$ at $\ell_A=3,6,10,13$, as shown in
Fig.~\ref{fig:mechanism}(b). The dominant $D=4$ component therefore
supports the peak structure without reproducing every cut exactly.

\subsubsection{Strong interaction}
Figs.~\ref{fig:amplificationN16}(g)--(i) present the strong-interaction
results. Unlike the nearly uniform ground state, the retained excited
state at $U/t=50$ has an asymmetric density and two second-R\'enyi
maxima at the edge cuts $\ell_A=1,15$.
At unit filling, $D=1$ contains exactly one doublon, one holon, and
fourteen singly occupied sites. The $D=1$ projected entanglement
$S_2^{\RR|1}$, shown in Fig.~\ref{fig:mechanism}(c), retains both
maxima.
This close agreement supports a predominantly
one-doublon--one-holon origin of the strong-coupling entropy profile.
The doublon is biased toward the left and
the holon toward the right, but neither has a fixed position.
When the cut passes through these boundary-weighted distributions,
different subsystem particle numbers compete. The edge entropy is
close to $\ln2$, consistent with nearly binary local occupation
uncertainty. A fixed, spatially separated doublon--holon Fock
configuration would instead be a product state with zero entanglement;
the relevant ingredient is the coherent spread of the carrier
positions across the cut.

\section{Skin modes and entanglement for other states}
\begin{figure*}[!tp]
 \includegraphics[width=\textwidth]{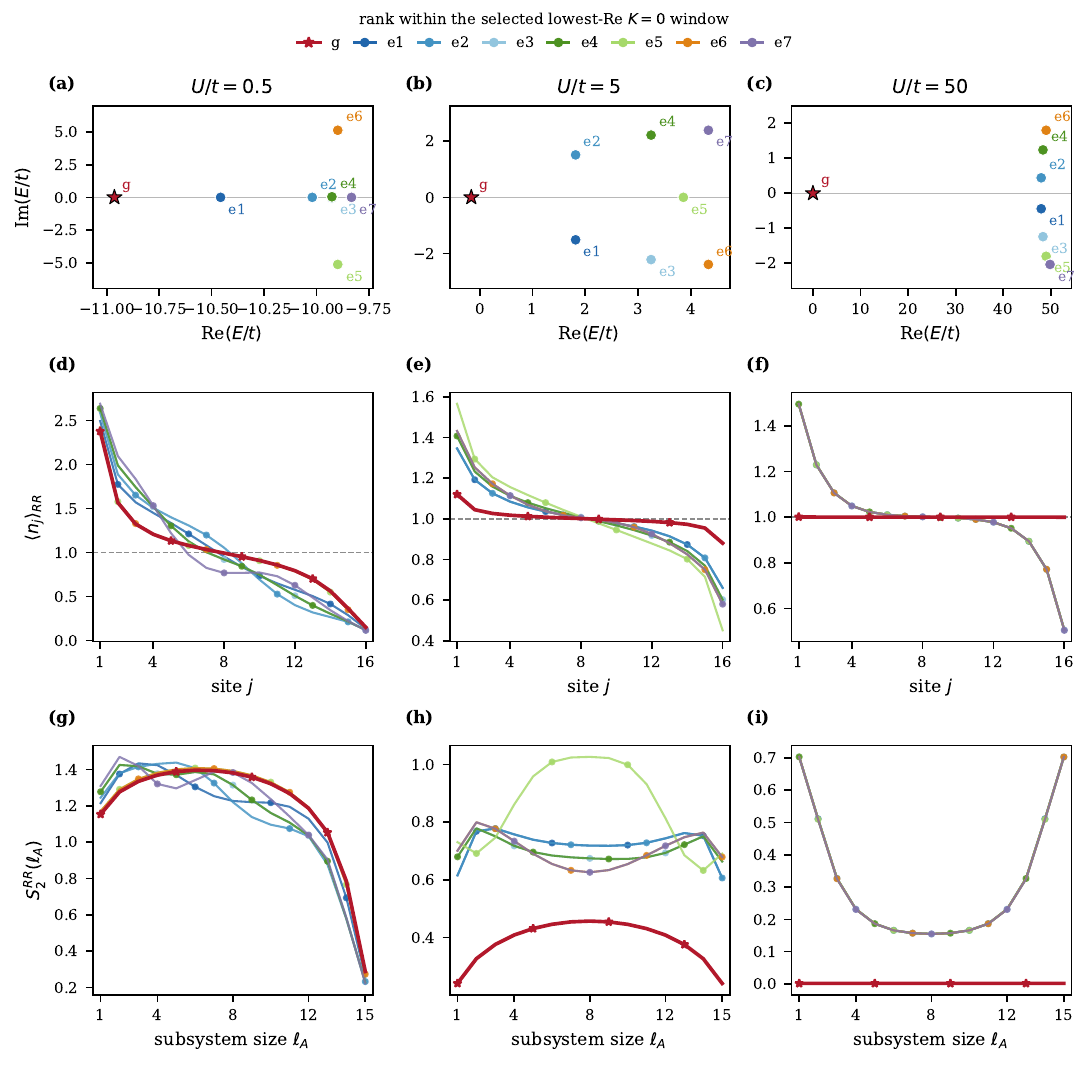}
 \caption{Selected low-real-energy states for $N=M=16$ and $\epsilon=0.1$;
 columns correspond to $U/t=0.5,5,50$.  (a)--(c) Complex energies of the
 minimum-$\operatorname{Re}E$ state and seven next-lowest-
 $\operatorname{Re}E$ states in $K=0$.  (d)--(f) Their normalized
 right-state densities.  (g)--(i) Their second R\'enyi entropies for
 every cut $\ell_A=1,\ldots,15$. }
 \label{fig:reproduction}
\end{figure*}

For completeness, we compare the density and second
R\'enyi entropy of eight selected low-$\operatorname{Re}E$ states
in $K=0$ at $U/t=0.5,5,50$ as shown in Fig.~\ref{fig:reproduction}.

At weak coupling, the ground state and selected
excitations show pronounced weak-link asymmetry.  At intermediate coupling (U/t=5), the
ground-state density approaches unit filling and its second Ren\'yi entropy is strongly
reduced.  Nearby excitations do not collapse onto one pattern: their
densities and complete cut profiles remain state selective.  
At $U/t=50$, these selected excited states show similar density and
entropy profiles, distinct from the ground-state curve.

\section{Conclusion}

In this paper, we studied skin modes, entanglement, and many-body winding
in the unidirectional Bose--Hubbard model. The large-system Bethe roots
provide a ground-state reference for the interaction-driven
superfluid--Mott transition. Flux continuation of these roots also
distinguishes a ground-connected energy that changes branch after one
flux quantum from a small closed loop on the Mott side, without
equating this single-state response with the full determinant winding.
 
 We compute the many-body winding number for different interaction strengths and discuss how the
suppression of skin modes relates to the many-body winding number. 
For the complete eight-site systems, the many-body winding about the origin remains nonzero at strong
interaction. The exact two-site example demonstrates that a nonzero
winding can persist while the corresponding spectral loop becomes
arbitrarily small. 
In this strongly interacting regime, 
the ground-state density is flat and the skin effect is strongly suppressed.

For sixteen sites, we found that the
ground-state density approaches unit filling and its entanglement is
strongly reduced. In contrast, the steady states
exhibit a nonmonotonic entanglement crossover. The ground-state
suppression of skin modes therefore does not describe the entanglement
of a different spectral selection.

We also showed that the four second-R\'enyi maxima at intermediate
interaction arise from the redistribution of subsystem particle-number
probabilities. Their variance provides a useful comparison but does not
reproduce the complete peak structure. The dominant global $D=4$
onsite-pair manifold retains all four maxima in its projected
second Ren\'yi entanglement. At $U/t=50$, the
$D=1$ projected entanglement closely reproduces the full second Ren\'yi entropy profile and
both edge maxima. This supports a one-doublon--one-holon picture
in which coherent positional uncertainty, rather than a fixed
separated pair, produces the entanglement.

These results show why skin density, entanglement, and point-gap topology
should be examined separately in an interacting non-Hermitian system.
It remains important to determine whether the observed crossover persists
under size scaling and in long-time evolution. This requires a consistent
state selection across sizes and a search that establishes the globally
dominant imaginary energy. We leave this for future work.

\begin{acknowledgments}
The authors used generative-AI tools (ChatGPT Sol
5.6, Astra 6,  and Claude Opus 5) for language assistance, code generation and  debugging,
numerical-workflow orchestration. The authors reviewed
and verified the derivations, numerical values, citations,
and final manuscript text and assumes full responsibility for the work. 
C.-F. L. thanks Mingchen Zheng for insightful discussion. 
P.-Y. C and C.-F. L. are supported by National Science and Technology Council of Taiwan under Grants No.
NSTC 114-2918-I-007-015, 115-2112-M-007-039. 
Y.-C. T is supported by National Science and Technology Council of Taiwan under Grants No.
NSTC 113-2112-M-A49-015-MY3.
The authors thank the support from the National Center for
Theoretical Sciences, Physics Division.
\end{acknowledgments}

\section*{Data and code availability}
The data that support the findings of this article are not
publicly available upon publication because it is not technically feasible and/or the cost of preparing, depositing, and
hosting the data would be prohibitive within the terms of this
research project. The data are available from the authors upon
reasonable request.

\bibliography{references}

@misc{chou2026,
      title={PT symmetry-enriched non-unitary criticality}, 
      author={Kuang-Hung Chou and Xue-Jia Yu and Po-Yao Chang},
      year={2025},
      eprint={2509.09587},
      archivePrefix={arXiv},
      primaryClass={quant-ph},
      url={https://arxiv.org/abs/2509.09587}, 
}

@article{Li2026,
  title = {Impurity-induced nonunitary criticality},
  author = {Li, Heng-Hsi and Chou, Kuang-Hung and Wen, Xueda and Chang, Po-Yao},
  journal = {Phys. Rev. B},
  volume = {113},
  issue = {3},
  pages = {035130},
  numpages = {5},
  year = {2026},
  month = {Jan},
  publisher = {American Physical Society},
  doi = {10.1103/m7lb-vkbf},
  url = {https://link.aps.org/doi/10.1103/m7lb-vkbf}
}

@article{Hsieh2023,
	title = {Relating non-Hermitian and Hermitian quantum systems at criticality},
	pages = {062},
	author = {Hsieh, Chang-Tse and Chang, Po-Yao},
	journal = {SciPost Phys. Core},
	volume = {6},
	year = {2023},
	publisher = {SciPost},
	doi = {10.21468/SciPostPhysCore.6.3.062},
	url = {https://scipost.org/10.21468/SciPostPhysCore.6.3.062}
}

@article{Martinez2018,
  title = {Non-Hermitian robust edge states in one dimension: Anomalous localization and eigenspace condensation at exceptional points},
  author = {Martinez Alvarez, V. M. and Barrios Vargas, J. E. and Foa Torres, L. E. F.},
  journal = {Phys. Rev. B},
  volume = {97},
  issue = {12},
  pages = {121401(R)},
  numpages = {6},
  year = {2018},
  month = {Mar},
  publisher = {American Physical Society},
  doi = {10.1103/PhysRevB.97.121401},
  url = {https://link.aps.org/doi/10.1103/PhysRevB.97.121401}
}

@article{Zheng2024,
  author = {Zheng, Mingchen and Qiao, Yi and Wang, Yupeng and Cao, Junpeng and Chen, Shu},
  title = {Exact Solution of the {Bose-Hubbard} Model with Unidirectional Hopping},
  journal = {Phys. Rev. Lett.},
  volume = {132},
  pages = {086502},
  year = {2024},
  doi = {10.1103/PhysRevLett.132.086502}
}

@article{Bergholtz2021,
  author = {Bergholtz, Emil J. and Budich, Jan Carl and Kunst, Flore K.},
  title = {Exceptional topology of non-Hermitian systems},
  journal = {Rev. Mod. Phys.},
  volume = {93},
  pages = {015005},
  year = {2021},
  doi = {10.1103/RevModPhys.93.015005}
}

@article{Gong2018,
  author = {Gong, Zongping and Ashida, Yuto and Kawabata, Kohei and Takasan, Kazuaki and Higashikawa, Sho and Ueda, Masahito},
  title = {Topological Phases of Non-Hermitian Systems},
  journal = {Phys. Rev. X},
  volume = {8},
  pages = {031079},
  year = {2018},
  doi = {10.1103/PhysRevX.8.031079}
}

@article{Okuma2020,
  author = {Okuma, Nobuyuki and Kawabata, Kohei and Shiozaki, Ken and Sato, Masatoshi},
  title = {Topological Origin of Non-Hermitian Skin Effects},
  journal = {Phys. Rev. Lett.},
  volume = {124},
  pages = {086801},
  year = {2020},
  doi = {10.1103/PhysRevLett.124.086801}
}

@article{Zhang2020,
  author = {Zhang, Kai and Yang, Zhesen and Fang, Chen},
  title = {Correspondence between Winding Numbers and Skin Modes in Non-Hermitian Systems},
  journal = {Phys. Rev. Lett.},
  volume = {125},
  pages = {126402},
  year = {2020},
  doi = {10.1103/PhysRevLett.125.126402}
}

@article{Kawabata2022,
  author = {Kawabata, Kohei and Shiozaki, Ken and Ryu, Shinsei},
  title = {Many-body topology of non-Hermitian systems},
  journal = {Phys. Rev. B},
  volume = {105},
  pages = {165137},
  year = {2022},
  doi = {10.1103/PhysRevB.105.165137}
}

@article{Mao2023,
  author = {Mao, Liang and Hao, Yajiang and Pan, Lei},
  title = {Non-Hermitian skin effect in a one-dimensional interacting Bose gas},
  journal = {Phys. Rev. A},
  volume = {107},
  pages = {043315},
  year = {2023},
  doi = {10.1103/PhysRevA.107.043315}
}

@article{Yoshida2024,
  author = {Yoshida, Tsuneya and Zhang, Song-Bo and Neupert, Titus and Kawakami, Norio},
  title = {Non-Hermitian Mott Skin Effect},
  journal = {Phys. Rev. Lett.},
  volume = {133},
  pages = {076502},
  year = {2024},
  doi = {10.1103/PhysRevLett.133.076502}
}

@article{Hamanaka2025,
  author = {Hamanaka, Shu and Kawabata, Kohei},
  title = {Multifractality of the many-body non-Hermitian skin effect},
  journal = {Phys. Rev. B},
  volume = {111},
  pages = {035144},
  year = {2025},
  doi = {10.1103/PhysRevB.111.035144}
}

@article{Kawabata2023,
  author = {Kawabata, Kohei and Numasawa, Tokiro and Ryu, Shinsei},
  title = {Entanglement Phase Transition Induced by the Non-Hermitian Skin Effect},
  journal = {Phys. Rev. X},
  volume = {13},
  pages = {021007},
  year = {2023},
  doi = {10.1103/PhysRevX.13.021007}
}

@article{Fu2023,
  author = {Fu, Yongxu and Zhang, Yi},
  title = {Hybrid scale-free skin effect in non-Hermitian systems: A transfer matrix approach},
  journal = {Phys. Rev. B},
  volume = {108},
  pages = {205423},
  year = {2023},
  doi = {10.1103/PhysRevB.108.205423}
}

@article{Calabrese2004,
  author = {Calabrese, Pasquale and Cardy, John},
  title = {Entanglement entropy and quantum field theory},
  journal = {J. Stat. Mech.},
  volume = {2004},
  pages = {P06002},
  year = {2004},
  doi = {10.1088/1742-5468/2004/06/P06002}
}

@article{Laflorencie2016,
  author = {Laflorencie, Nicolas},
  title = {Quantum entanglement in condensed matter systems},
  journal = {Phys. Rep.},
  volume = {646},
  pages = {1},
  year = {2016},
  doi = {10.1016/j.physrep.2016.06.008}
}

@article{Fisher1989,
  author = {Fisher, Matthew P. A. and Weichman, Peter B. and Grinstein, G. and Fisher, Daniel S.},
  title = {Boson localization and the superfluid-insulator transition},
  journal = {Phys. Rev. B},
  volume = {40},
  pages = {546},
  year = {1989},
  doi = {10.1103/PhysRevB.40.546}
}

@article{Eisert2010,
 author={Eisert, J. and Cramer, M. and Plenio, M. B.},
 title={Colloquium: Area laws for the entanglement entropy},
 journal={Rev. Mod. Phys.}, volume={82}, pages={277--306}, year={2010},
 doi={10.1103/RevModPhys.82.277}
}

@article{Feng2024,
 author={Feng, Xu and Chen, Shu},
 title={Delocalization of skin steady states},
 journal={Phys. Rev. B}, volume={110}, pages={144305}, year={2024},
 doi={10.1103/PhysRevB.110.144305}
}

@article{Goldstein2018,
 author={Goldstein, Moshe and Sela, Eran},
 title={Symmetry-Resolved Entanglement in Many-Body Systems},
 journal={Phys. Rev. Lett.}, volume={120}, pages={200602}, year={2018},
 doi={10.1103/PhysRevLett.120.200602}
}

@article{Lukin2019,
 author={Lukin, Alexander and Rispoli, Matthew and Schittko, Robert and
 Tai, M. Eric and Kaufman, Adam M. and Choi, Soonwon and Khemani, Vedika and
 L{\'e}onard, Julian and Greiner, Markus},
 title={Probing entanglement in a many-body-localized system},
 journal={Science}, volume={364}, pages={256--260}, year={2019},
 doi={10.1126/science.aau0818}
}

@article{Hatano1996,
  author = {Hatano, Naomichi and Nelson, David R.},
  title = {Localization Transitions in Non-Hermitian Quantum Mechanics},
  journal = {Phys. Rev. Lett.},
  volume = {77},
  pages = {570--573},
  year = {1996},
  doi = {10.1103/PhysRevLett.77.570}
}

@article{Yao2018,
 author={Yao, Shunyu and Wang, Zhong},
 title={Edge States and Topological Invariants of Non-Hermitian Systems},
 journal={Phys. Rev. Lett.}, volume={121}, pages={086803}, year={2018},
 doi={10.1103/PhysRevLett.121.086803}
}

@article{Kunst2018,
 author={Kunst, Flore K. and Edvardsson, Elisabet and Budich, Jan Carl and Bergholtz, Emil J.},
 title={Biorthogonal Bulk-Boundary Correspondence in Non-Hermitian Systems},
 journal={Phys. Rev. Lett.}, volume={121}, pages={026808}, year={2018},
 doi={10.1103/PhysRevLett.121.026808}
}

@article{Yokomizo2019,
 author={Yokomizo, Kazuki and Murakami, Shuichi},
 title={Non-Bloch Band Theory of Non-Hermitian Systems},
 journal={Phys. Rev. Lett.}, volume={123}, pages={066404}, year={2019},
 doi={10.1103/PhysRevLett.123.066404}
}

@article{Herviou2019,
 author={Herviou, Lo{\"i}c and Regnault, Nicolas and Bardarson, Jens H.},
 title={Entanglement spectrum and symmetries in non-Hermitian fermionic non-interacting models},
 journal={SciPost Phys.}, volume={7}, pages={069}, year={2019},
 doi={10.21468/SciPostPhys.7.5.069}
}

@article{Chang2020,
 author={Chang, Po-Yao and You, Jhih-Shih and Wen, Xueda and Ryu, Shinsei},
 title={Entanglement spectrum and entropy in topological non-Hermitian systems and nonunitary conformal field theory},
 journal={Phys. Rev. Research}, volume={2}, pages={033069}, year={2020},
 doi={10.1103/PhysRevResearch.2.033069}
}

@article{Tu2022,
 author={Tu, Yi-Ting and Tzeng, Yu-Chin and Chang, Po-Yao},
 title={R{\'e}nyi entropies and negative central charges in non-Hermitian quantum systems},
 journal={SciPost Phys.}, volume={12}, pages={194}, year={2022},
 doi={10.21468/SciPostPhys.12.6.194}
}

@article{Lieb1963,
 author={Lieb, Elliott H. and Liniger, Werner},
 title={Exact Analysis of an Interacting {Bose} Gas. {I}. The General Solution and the Ground State},
 journal={Phys. Rev.}, volume={130}, pages={1605--1616}, year={1963},
 doi={10.1103/PhysRev.130.1605}
}

@article{Bogoliubov1993,
 author={Bogoliubov, N. M. and Bullough, R. K. and Pang, G. D.},
 title={Exact solution of a {$q$}-boson hopping model},
 journal={Phys. Rev. B}, volume={47}, pages={11495--11498}, year={1993},
 doi={10.1103/PhysRevB.47.11495}
}

@article{Links2003,
 author={Links, Jon and Zhou, Huan-Qiang and McKenzie, Ross H. and Gould, Mark D.},
 title={Algebraic {Bethe} ansatz method for the exact calculation of energy spectra and form factors: applications to models of {Bose--Einstein} condensates and metallic nanograins},
 journal={J. Phys. A: Math. Gen.}, volume={36}, pages={R63--R104}, year={2003},
 doi={10.1088/0305-4470/36/19/201}
}

@article{Links2006,
 author={Links, Jon and Hibberd, Katrina E.},
 title={{Bethe} Ansatz Solutions of the {Bose--Hubbard} Dimer},
 journal={SIGMA}, volume={2}, pages={095}, year={2006},
 doi={10.3842/SIGMA.2006.095}
}

@article{Choy1982,
 author={Choy, T. C. and Haldane, F. D. M.},
 title={Failure of {Bethe-Ansatz} solutions of generalisations of the {Hubbard} chain to arbitrary permutation symmetry},
 journal={Phys. Lett. A}, volume={90}, pages={83--84}, year={1982},
 doi={10.1016/0375-9601(82)90057-3}
}

@article{Albertini1996,
 author={Albertini, Giuseppe and Dahmen, Silvio Renato and Wehefritz, Birgit},
 title={Phase diagram of the non-Hermitian asymmetric {XXZ} spin chain},
 journal={J. Phys. A: Math. Gen.}, volume={29}, pages={L369--L376}, year={1996},
 doi={10.1088/0305-4470/29/15/001}
}

@book{Korepin1993,
 author={Korepin, V. E. and Bogoliubov, N. M. and Izergin, A. G.},
 title={Quantum Inverse Scattering Method and Correlation Functions},
 publisher={Cambridge University Press}, address={Cambridge}, year={1993},
 doi={10.1017/CBO9780511628832}
}

@article{Xavier2018,
 author={Xavier, J. C. and Alcaraz, F. C. and Sierra, G.},
 title={Equipartition of the entanglement entropy},
 journal={Phys. Rev. B}, volume={98}, pages={041106}, year={2018},
 doi={10.1103/PhysRevB.98.041106}
}

\end{document}